\documentclass[preprint2]{aastex}
\usepackage{color}
\usepackage{graphicx}
\usepackage{bm}
\usepackage{amsmath}

\begin{document}
\title{POLARIZED LINE FORMATION IN MULTI-DIMENSIONAL MEDIA. IV. 
A FOURIER DECOMPOSITION TECHNIQUE TO FORMULATE THE TRANSFER
EQUATION WITH ANGLE-DEPENDENT PARTIAL
FREQUENCY REDISTRIBUTION}
\author{L.~S.~Anusha$^{1}$ and K.~N.~Nagendra$^{1}$}
\affil{$^1$Indian Institute of Astrophysics, Koramangala,
2nd Block, Bangalore 560 034, India}

\begin{abstract}
To explain the linear polarization observed in spatially resolved structures
in the solar atmosphere, the solution of polarized radiative 
transfer (RT) equation in multi-dimensional (multi-D) geometries is essential.
For strong resonance lines partial frequency redistribution (PRD)
effects also become important. In a series of papers we have been investigating
the nature of Stokes profiles formed in multi-D media including 
PRD in line scattering. For numerical simplicity so far we restricted 
our attention to the particular case of PRD functions which are averaged over 
all the incident and scattered directions. In this paper we 
formulate the polarized RT equation in multi-D media
that takes into account Hanle effect with angle-dependent PRD functions. 
We generalize here to the multi-D case, the method of Fourier series 
expansion of angle-dependent PRD functions originally developed for 
RT in 1D geometry. We show that the Stokes source vector 
$\bm{S}=(S_I,S_Q,S_U)^T$ and the Stokes vector $\bm{I}=(I,Q,U)^T$ 
can be expanded in terms of infinite sets of components 
${\tilde{\bm{\mathcal{S}}}}^{(k)}$, ${\tilde{\bm{\mathcal{I}}}}^{(k)}$ 
respectively, $k\in[0,+\infty)$. 
We show that the components ${\tilde{\bm{\mathcal{S}}}}^{(k)}$ become 
independent of the azimuthal angle ($\varphi$) of the scattered ray, 
whereas the components
${\tilde{\bm{\mathcal{I}}}}^{(k)}$ remain dependent on $\varphi$ due to
the nature of RT in multi-D geometry. We also establish that
${\tilde{\bm{\mathcal{S}}}}^{(k)}$ and 
${\tilde{\bm{\mathcal{I}}}}^{(k)}$ satisfy a
simple transfer equation, which can be solved by any iterative method
like an Approximate Lambda Iteration (ALI) or a Biconjugate-Gradient type
projection method provided we truncate the Fourier series to have a finite
number of terms.
\end{abstract}


\section{INTRODUCTION}
\label{introduction}
The observations of the solar atmospheres reveal a
wealth of information about the spatially inhomogeneous 
structures. Modern spectro-polarimeters with high spatial and
polarimetric resolution are able to distinguish the changes
in the linearly polarized spectrum caused by such structures.
To model the spectro-polarimetric observations 
of such spatially resolved structures, one has to solve a three dimensional 
(3D) polarized line radiative transfer (RT) equation. A historical 
account of the developments of RT in multi-dimensional (multi-D) media 
is presented in detail in \citet[][hereafter called Paper I]{anuknn11a}. 
In a series of papers
we have been investigating the nature of linearly 
polarized profiles formed in multi-D  media taking 
account of the partial frequency redistribution (PRD) in line scattering. 
In  Paper I we developed a method of `Stokes
vector decomposition' in multi-D geometry 
in terms of `irreducible spherical 
tensors' ${\mathcal T}^K_Q$ \citep[see][]{ll04}. 
It was a generalization to the multi-D case, of the decomposition technique
developed in \citet[][hereafter HF07]{hf07} for the one-dimensional (1D) case.
In \citet[][hereafter Paper II]{anuetal11b}, we developed a fast
numerical method called the Pre-BiCG-STAB (Stabilized preconditioned 
Bi-Conjugate Gradient), to solve the polarized RT problems in two dimensional 
(2D) media. In \citet[][hereafter Paper III]{anuknn11c} we generalized 
the works of Papers I and II to include scattering in the presence 
of weak magnetic fields (Hanle effect) in a 3D geometry.
In all these papers we considered only angle-averaged PRD functions. 

The polarized Stokes line transfer problems with angle-dependent PRD 
in 1D planar geometries were solved by several authors
\citep[see][]{dumetal77,mck85,mf87,mf88,knnetal02,knnetal03,sametal08}.
In this formalism, a strong
coupling of incident and scattered ray directions 
($\bm{\Omega'}$ and $\bm{\Omega}$ respectively) prevails in 
the scattering phase matrices
as well as the angle-dependent PRD functions, which brings
in unmitigated numerical difficulties. To simplify the problem,
a method based on `decomposition of the phase matrices' in terms of 
${\mathcal T}^K_Q$ combined with a `Fourier
series expansion' of the angle-dependent redistribution functions
$r_{\rm II, III}(x,x',\bm{\Omega},\bm{\Omega'})$ of \citet[][]{hum62} 
was proposed for Hanle and Rayleigh scattering by 
\citet[][hereafter HF09]{hf09} and
\citet[][]{hf10} respectively. \citet{sametal11a} developed efficient numerical
methods to solve angle-dependent RT problems 
for the case of Rayleigh scattering, based on the decomposition 
technique developed in \citet[][]{hf10}. \citet{sam11b} 
proposed a single scattering approximation 
to solve the more difficult problem of RT with 
angle-dependent PRD including
Hanle effect. However all these works are confined to the limit of 
1D planar geometry.

In this paper we generalize to the multi-D case, the Fourier decomposition
technique developed in HF09 for the 1D case. In the first step we decompose
the phase matrices in terms of ${\mathcal T}^K_Q$ as done in Paper I and
III. However we now formulate a polarized RT equation for multi-D that also 
includes angle-dependent PRD functions. 
We set up a transfer equation in terms of a new set of 6-dimensional 
vectors called as `irreducible source and the irreducible
Stokes vectors'. 
In the second step, we expand the $r_{\rm II, III}(x,x',\bm{\Omega},
\bm{\Omega'})$ redistribution functions in terms of a 
Fourier series with respect to the azimuthal angle $(\varphi)$ of the 
scattered ray. 
Then we transform the original RT equation into a new RT equation
which is simpler to solve because the latter has smaller number
of independent variables.
This simplified (reduced) transfer equation 
can be solved by any iterative method
like the Approximate Lambda Iteration (ALI) or a Bi-Conjugate Gradient type
projection method.

In Table~\ref{table_1} we list the important
milestones in the specific area of `formulation and solution of polarized
RT equation' with resonance scattering and/or Hanle effect in 1D
and multi-D media in different formalisms. 
The emphasis is on showing how the complexity of the problem is reduced to
manageable levels, by the concerted efforts of several authors.
It includes a brief
historical account of the formulation and decomposition
of polarized phase matrices and the redistribution matrices
for spectral lines. 
In the literature on this topic, the term `phase matrix' refers
only to the angular correlations in the polarized light scattering
\citep[see e.g., the Rayleigh scattering polarized phase matrix 
described in][]{chandra60}. The phase matrices are in general
frequency independent. The `redistribution matrix' on
the other hand, contains both frequency and angle correlations 
between the incident and scattered
photons. The formulation of the redistribution matrices 
in the astrophysical
literature (in the modern analytic form), dates back
to the pioneering work of \citet{osc72,osc73}. 
The references given here serve only to mark the milestones.
No pretension is made to give a full list of references. 

In Section~\ref{rte} we describe the multi-D transfer
equation in the Stokes vector formalism. An irreducible 
transfer equation for angle-dependent PRD functions in
multi-D media is presented in Section~\ref{reduced-rte}. In
Section~\ref{fourier-rte} a transfer equation in multi-D geometry 
for the irreducible Fourier coefficients of the Stokes source vector
and the Stokes vector is established.
Conclusions are given in Section~\ref{conclusions}.

\section{TRANSFER EQUATION IN TERMS OF STOKES PARAMETERS}
\label{rte}
\begin{figure*}
\centering
\includegraphics[scale=0.5]{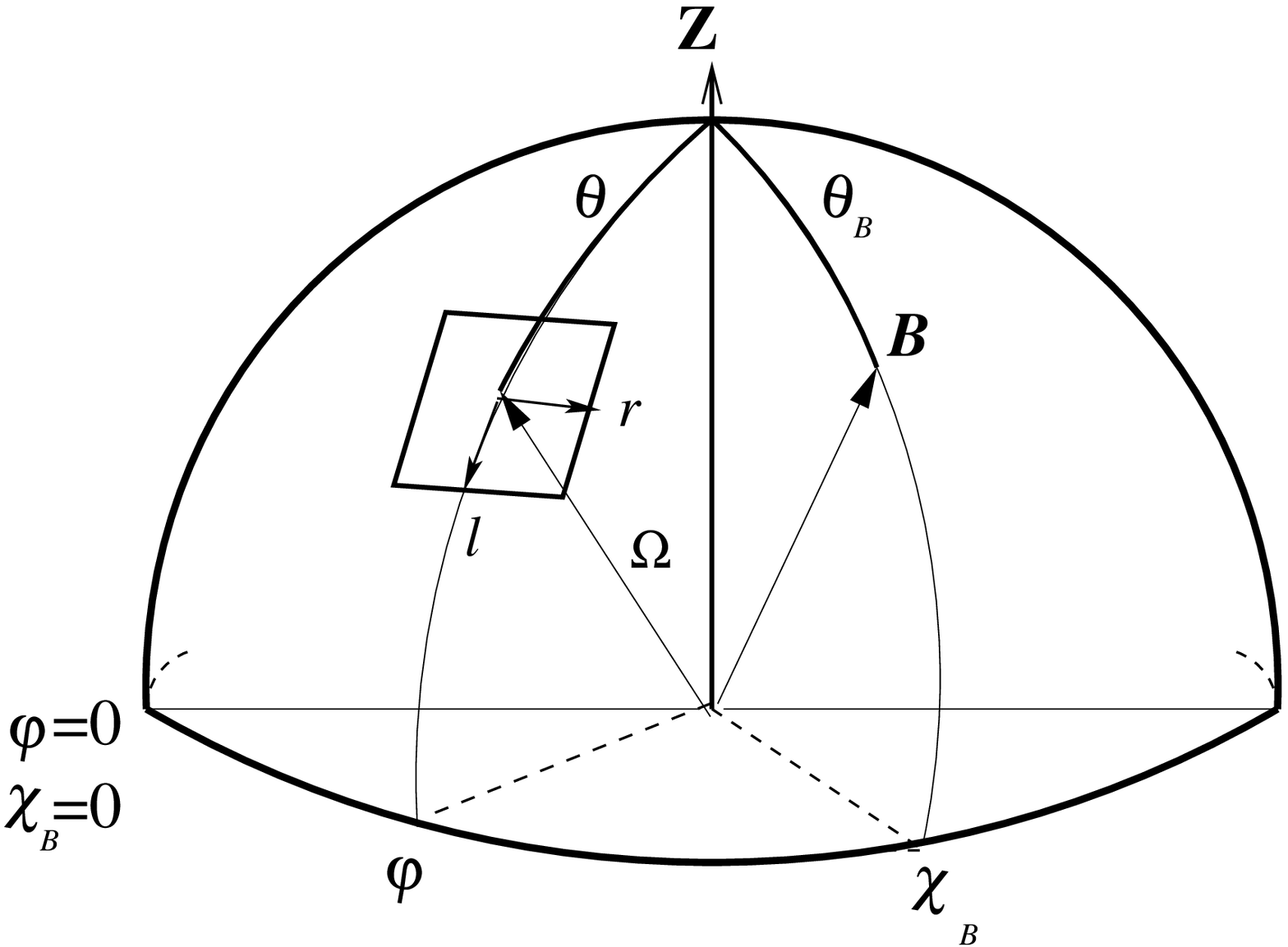}
\includegraphics[scale=0.35]{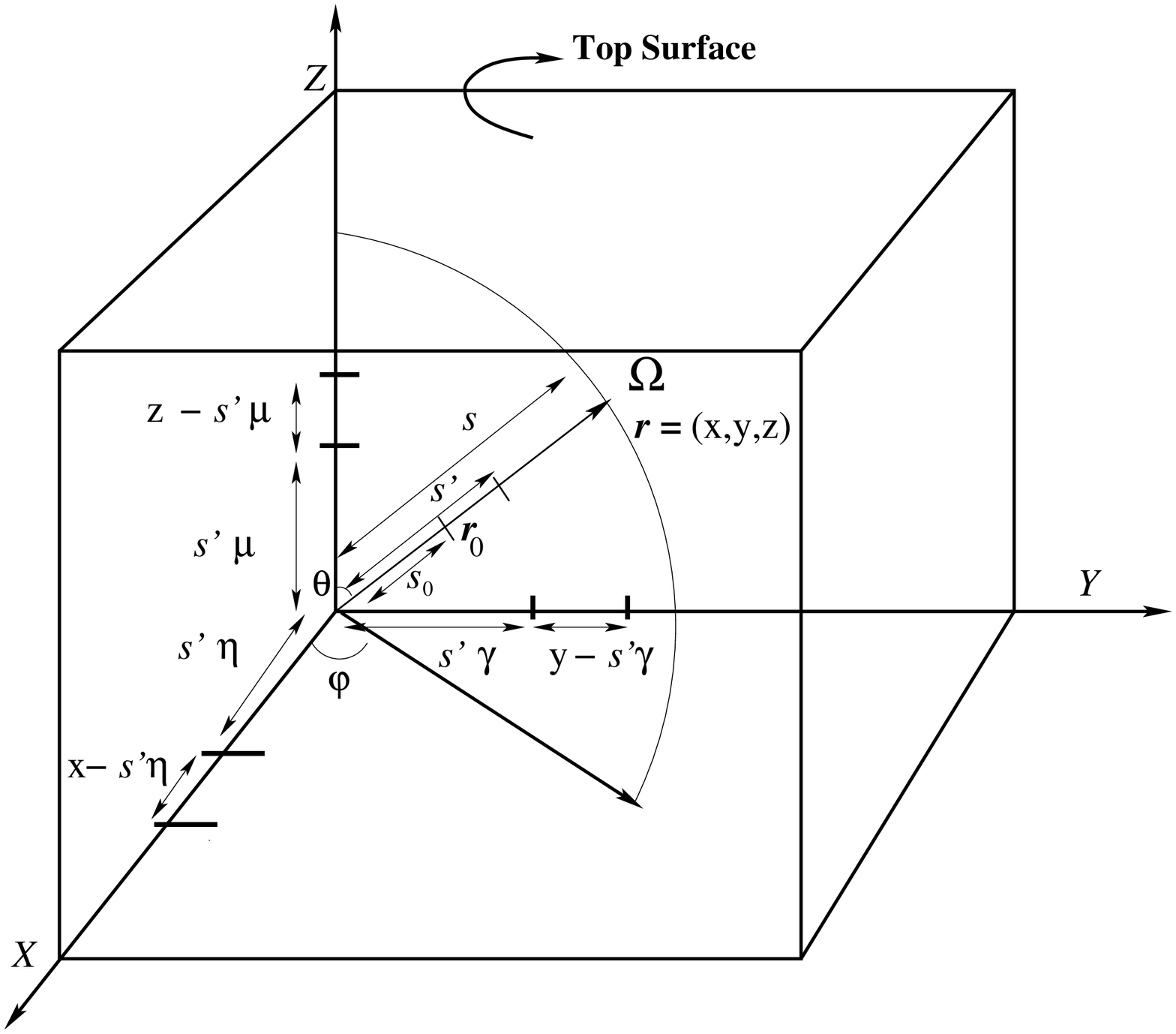}
\caption{In the top panels we show atmospheric reference frame.
The angle pair $(\theta,\varphi)$ define
the scattered ray direction and $(\theta',\varphi')$ that
of the incident ray direction. The magnetic field is
characterized by $\bm{B}=(\Gamma,\theta_B,\chi_B)$, where $\Gamma$ is
the Hanle efficiency parameter and ($\theta_B,\chi_B$)
define the field direction. In the bottom panels we show the
definition of the position vector $\bm{r}$ and the
projected distances $\bm{r}-s'\bm{\Omega}$ which appear in
Equation~(\ref{3d-formal}). $\bm{r}_0$ and $\bm{r}$ are the
arbitrary initial and final locations that appear in the formal
solution integral (Equation~(\ref{3d-formal})).
}
\label{fig-scatgeo-fs}
\end{figure*}

For a given ray defined by the direction $\bm{\Omega}$, the 
polarized transfer equation in a multi-D medium 
for a two-level model atom with unpolarized ground level is given by 
\begin{eqnarray}
\!\!\!\!\!\!&&\bm{\Omega} \cdot \bm{\nabla}\bm{I}(\bm{r}, \bm{\Omega}, x) 
= -[\kappa_l(\bm{r}) \phi(x)+\kappa_c(\bm{r})]
\nonumber \\
\!\!\!\!\!\!&& \times
[\bm{I}(\bm{r}, \bm{\Omega}, x)-\bm{S}(\bm{r}, 
\bm{\Omega}, x)].
\label{3d-rte}
\end{eqnarray}
Analogous equations such as Equation~(\ref{3d-rte})
for the unpolarized case can be found in several references 
\citep[see e.g.,][]{adam90,mam78,pom73}. For the polarized 
case with PRD, the transfer equations are given in Papers I, II and III.
Here $\bm{I}=(I,Q,U)^T$ is the Stokes vector with
$I$, $Q$ and $U$ the Stokes parameters defined as in 
\citet{chandra60}. 
The reference directions $l$ and $r$ are marked in the
top panels of Figure~\ref{fig-scatgeo-fs}. Positive value
of $Q$ is defined to be in a direction parallel to $l$
and negative $Q$ in a direction parallel to $r$. 
The quantity $\bm{r}=({\rm x},{\rm y}, 
{\rm z})$ is the position vector of the ray in the 
Cartesian co-ordinate system (see bottom panels of Figure~\ref{fig-scatgeo-fs}). 
The unit vector $\bm{\Omega}=(\eta,\gamma,\mu)=
(\sin\theta\,\cos \varphi\,,\sin\theta\,\sin \varphi\,,\cos \theta)$ 
defines the direction cosines of the ray in the atmosphere with respect to
the atmospheric normal (the $Z$-axis), where $\theta$ and 
$\varphi$ are the polar 
and azimuthal angles of the ray (see Figure~\ref{fig-scatgeo-fs}).

The quantity $\kappa_l$ is the
frequency averaged line opacity,
$\phi$ is the Voigt profile function and $\kappa_c$ is the continuum opacity.
Frequency is measured in reduced units, namely
$x=(\nu-\nu_0)/\Delta \nu_D$ where $\Delta \nu_D$ is the Doppler width.
The Stokes source vector in a two-level model atom with unpolarized 
ground level \citep[see e.g.,][]{mf87,knnetal02} is  
\begin{eqnarray}
&&\bm{S}(\bm{r}, \bm{\Omega}, x)\nonumber \\
&&=\frac{\kappa_l(\bm{r}) \phi(x)
\bm{S}_{l}(\bm{r}, \bm{\Omega}, x)+\kappa_c(\bm{r})
\bm{S}_{c}(\bm{r}, x)}
{\kappa_l(\bm{r}) \phi(x)+\kappa_c(\bm{r})}. \nonumber \\
\label{s-tot}
\end{eqnarray}
$\bm{S}_{c}$ is the unpolarized continuum source vector given by
$(B_{\nu}(\bm{r}), 0, 0)^T$ with $B_{\nu}(\bm{r})$ being the 
Planck function. The line source vector 
\citep[see e.g.,][]{mf87,knnetal02} is written as  
\begin{eqnarray}
&&\!\!\!\!\!\!\!\!\!\!\!\!\!\!\!\!\!\!
\bm{S}_{l}(\bm{r}, \bm{\Omega}, x)
=\bm{G}(\bm{r})+\int_{-\infty}^{+\infty}dx' \nonumber \\
&&\!\!\!\!\!\!\!\!\!\!\!\!\!\!\!\!\!\!
\times \oint\frac{d\bm{\Omega}'}{4 \pi} 
\frac{{\hat{R}}(x, x', \bm{\Omega}, \bm{\Omega}', \bm{B})} {\phi(x)} 
\bm{I}(\bm{r}, \bm{\Omega}', x'). \nonumber \\
\label{s-line}
\end{eqnarray}
$\hat {R}$ is the Hanle redistribution matrix with
angle dependent PRD \citep[see Section~4.2, 
approximation level II of][]{bom97b}.
$\bm{B}$ represents an oriented vector magnetic field.
The thermalization parameter $\epsilon=\Gamma_I/(\Gamma_R+\Gamma_I)$
with $\Gamma_I$ and $\Gamma_R$ being the inelastic collision rate and the
radiative de-excitation rate respectively. 
The damping parameter is computed using $a=a_R [1+(\Gamma_E+\Gamma_I)/\Gamma_R]$
where $a_R={\Gamma_R}/{4 \pi \Delta \nu_D}$ and $\Gamma_E$ is the
elastic collision rate.
We denote the thermal source vector by 
$\bm{G}(\bm{r})=\epsilon \bm{B}_{\nu}(\bm{r})$
with $\bm{B}_{\nu}(\bm{r})=(B_{\nu}(\bm{r}), 0, 0)^T$.
The solid angle
element $d \bm{\Omega}'=\sin \theta'\,d \theta'\,d 
\varphi'$ where $\theta \in [0, \pi]$ and $\varphi \in [0, 2\pi]$.
The transfer equation along the ray path takes the form
\begin{eqnarray}
&&\!\!\!\!\!\!\!\!\!\!\!\!\!\!\!\!\!\!
\frac{d\bm{I}(\bm{r}, \bm{\Omega}, x)}{ds}\nonumber \\
&&\!\!\!\!\!\!\!\!\!\!\!\!\!\!\!\!\!\!
=-\kappa_{\rm tot}(\bm{r}, x)
[\bm{I}(\bm{r}, \bm{\Omega}, x)-\bm{S}(\bm{r}, 
\bm{\Omega}, x)],\nonumber \\
\label{3d-rte-path}
\end{eqnarray}
where $s$ is the path length along the ray and 
$\kappa_{\rm tot}(\bm{r}, x)$ is the
total opacity given by
\begin{equation}
\kappa_{tot}(\bm{r}, x)=\kappa_l(\bm{r}) \phi(x)+\kappa_c(\bm{r}).
\label{ktot}
\end{equation}

The formal solution of Equation~(\ref{3d-rte-path}) is given by
\begin{eqnarray}
&&\!\!\!\!\!\!\!\!\bm{I}(\bm{r}, \bm{\Omega}, x)\nonumber\\
&&\!\!\!\!\!\!\!\!=\bm{I}(\bm{r}_0, \bm{\Omega}, x)
\exp\left\{-{\int_{s_0}^{s}} \kappa_{\rm tot}(\bm{r}-s^{\prime\prime}
\bm{\Omega}, x) ds^{\prime\prime}\right\}\nonumber \\
&&+\int_{s_0}^{s} \bm{S}(\bm{r}
-s^{\prime}\bm{\Omega}, \bm{\Omega}, x)
\kappa_{\rm tot}(\bm{r}-s^{\prime}\bm{\Omega},x)\nonumber\\
&&\times\exp\left\{-\displaystyle{\int_{s'}^{s}}  
\kappa_{\rm tot}(\bm{r}-s^{\prime\prime}
\bm{\Omega},x) ds^{\prime\prime}\right\}
ds^{\prime}.
\label{3d-formal}
\end{eqnarray}
$\bm{I}(\bm{r}_0, \bm{\Omega}, x)$ is the boundary condition
imposed at $\bm{r}_0=({\rm x}_0,{\rm y}_0,{\rm z}_0)$.
Here $s$ is the distance measured along the ray path 
(see bottom panels of Figure~\ref{fig-scatgeo-fs}). Equations~(\ref{3d-rte})--(\ref{3d-formal}) 
can be solved using a perturbation method 
\citep[see][for the corresponding 1D case]{knnetal02}. 
However the perturbation method involves 
an approximation that the degree of linear polarization
is small (few \% only). Under the situations where the degree of polarization
becomes large the perturbation method cannot be expected to guarantee
a stable solution. A numerical disadvantage of working in Stokes vector
formalism is that the physical quantities depend on 
all the angular variables ($\bm{\Omega}, \bm{\Omega'}$). Added
to this, the angle-dependent polarized RT problem demands high
angular grid resolution thereby requiring enormous memory and
CPU time.

\section{TRANSFER EQUATION IN TERMS OF IRREDUCIBLE SPHERICAL TENSORS}
\label{reduced-rte}
As shown in HF07, $\bm{S}$ and $\bm{I}$
can be decomposed into 6-dimensional cylindrically
symmetrical vectors $\bm{\mathcal S}$ and $\bm{\mathcal I}$
defined for a 1D geometry as 
\begin{eqnarray}
&&\bm{\mathcal S}=(S^0_0,
S^2_0,S^{2,{\rm x}}_1, S^{2,{\rm y}}_1,
S^{2,{\rm x}}_2, S^{2,{\rm y}}_2)^T,\nonumber\\
&&\bm{\mathcal I}=(I^0_0,
I^2_0, I^{2,{\rm x}}_1, I^{2,{\rm y}}_1,
I^{2,{\rm x}}_2, I^{2,{\rm y}}_2)^T.
\end{eqnarray}
In Papers I and III generalizations of the technique
of HF07 to the multi-D case are discussed, for the
case of angle-averaged PRD. 
We show here that the same decomposition method can be applied to 
the corresponding angle-dependent PRD case, by replacing the
angle-averaged PRD functions with angle-dependent PRD functions. 
This leads to an additional dependence of
$\bm{\mathcal S}$ on the scattered ray direction $\bm{\Omega}$.
The vectors $\bm{\mathcal I}$ and $\bm{\mathcal S}$ satisfy a transfer
equation of the form
\begin{eqnarray}
&&-\frac{1}{\kappa_{\rm tot}(\bm{r}, x)}\bm{\Omega} \cdot
\bm{\nabla}\bm{\mathcal{I}}(\bm{r}, \bm{\Omega}, x) = \nonumber \\
\!\!\!\!\!\!&&[\bm{\mathcal{I}}(\bm{r}, \bm{\Omega}, x)-
\bm{\mathcal S}(\bm{r}, \bm{\Omega}, x)],
\label{rte-reduced}
\end{eqnarray}
where 
\begin{equation}
\bm{\mathcal S}(\bm{r}, \bm{\Omega}, x)=p_x 
\bm{\mathcal S}_{l}(\bm{r}, \bm{\Omega}, x)
+(1-p_x)\bm{\mathcal S}_C(\bm{r}, x),
\label{stot-reduced}
\end{equation}
and 
\begin{equation}
p_x=\kappa_l(\bm{r}) \phi(x) / \kappa_{\rm tot}(\bm{r}, x).
\label{px}
\end{equation}
The irreducible line source vector is given by
\begin{eqnarray}
\!\!\!\!\!\!&&\bm{\mathcal S}_{l}(\bm{r}, \bm{\Omega}, x)=
\bm{G}(\bm{r})+\frac{1}{\phi(x)} \int_{-\infty}^{+\infty} dx' \nonumber \\
\!\!\!\!\!\!&& \times
\oint\frac{d \bm{\Omega}'} {4 \pi} 
\hat{W}\Big\{\hat{M}_{\rm II}(\bm{B},x,x')
r_{\rm II}(x, x', \bm{\Omega}, \bm{\Omega'}) \nonumber \\
\!\!\!\!\!\!&&+\hat{M}_{\rm III}(\bm{B},x,x')
r_{\rm III}(x, x',\bm{\Omega}, \bm{\Omega'}) \Big\}
\hat{\Psi}(\bm{\Omega}') \nonumber \\
&&\times \bm{\mathcal{I}}(\bm{r}, \bm{\Omega}',x'),
\label{sl-reduced}
\end{eqnarray}
with $\bm{G}(\bm{r})=(\epsilon \bm{B}_{\nu}(\bm{r}),0,0,0,0,0)^T$ and 
the irreducible unpolarized continuum source vector 
$\bm{\mathcal S}_C(\bm{r}, x)$
=$(S_C(\bm{r},x),0,0,0,0,0)^T$.
We assume that $S_C(\bm{r}, x)=B_{\nu}(\bm{r})$. 
$\hat{W}$ is a diagonal matrix written as
\begin{equation}
\hat{W}=\textrm{diag}\{W_0,W_2,W_2,W_2,W_2,W_2\}.
\label{w}
\end{equation}
$r_{\rm II,III}$ are the well known angle-dependent PRD
functions of \citet[][]{hum62} which depend explicitly
on the scattering angle $\Theta$, defined through 
$\cos \Theta = \bm{\Omega} \cdot \bm{\Omega'}$ computed using
\begin{equation}
\cos \Theta = \mu \mu' + \sqrt{(1-\mu^2)(1-\mu'^2}) \cos (\varphi'-\varphi).
\label{captheta}
\end{equation}
The matrix $\hat{\Psi}$ represents the reduced phase matrix for
the Rayleigh scattering. Its elements are listed in Appendix D 
of Paper III. The elements of the 
matrices $\hat{M}_{\rm II,III}(\bm{B},x,x')$ 
can be found in \citet{bom97b}.
The formal solution now takes the form
\begin{eqnarray}
&&\!\!\!\!\!\!\!\!\!\bm{\mathcal{I}}(\bm{r},\bm{\Omega}, x)=
\bm{\mathcal {I}}(\bm{r}_0, \bm{\Omega}, x)
e^{-\tau_{x, max}}\nonumber \\
&&\!\!\!\!\!\!\!\!\!
+\int_{0}^{\tau_{x, max}} 
e^{-\tau'_{x}(\bm{r}')}
\bm{\mathcal {S}}(\bm{r}',  \bm{\Omega}, x)d\tau'_{x}(\bm{r}').
\label{i-out-tau}
\end{eqnarray}
Here $\bm{\mathcal {I}}(\bm{r}_0, \bm{\Omega}, x)$ 
is the boundary condition imposed at $\bm{r}_0$.
The monochromatic optical depth scale is defined as
\begin{equation}
\tau_x({\rm x},{\rm y}, {\rm z})=\int_{s_0}^s \kappa_{\rm tot}(\bm{r}-s''
\bm{\Omega}, x)\, ds'',
\label{tau}
\end{equation}
where $\tau_x$ is measured along
a given ray determined by the direction $\bm{\Omega}$.
In Equation~(\ref{i-out-tau}) $\tau_{x, max}$ is the maximum
monochromatic optical depth at frequency $x$ when measured along the ray.

One can develop iterative methods to solve
Equations~(\ref{rte-reduced})--(\ref{i-out-tau}).
Because the physical quantities (like $\bm{\mathcal {S}}$) still
depend on $\bm{\Omega}$, it is not numerically very efficient.
In the next section we present a method to transform the Equation~(\ref{rte-reduced})
into a new RT equation, which is simpler to solve.
\section{TRANSFER EQUATION IN TERMS OF IRREDUCIBLE FOURIER COEFFICIENTS} 
\label{fourier-rte}
HF09 introduced a method of Fourier series expansion of the
angle-dependent PRD functions 
$r_{\rm II,III}(x,x',\bm{\Omega},\bm{\Omega'})$. 
Here we present a generalization to the multi-D case, the
formulation given in HF09. 

\noindent
{\bf Theorem:} In a multi-D polarized RT including angle-dependent PRD and 
Hanle effect, the irreducible source vector $\bm{\mathcal {S}}$ and the
irreducible Stokes vector $\bm{\mathcal {I}}$ exhibit Fourier expansions
of the form
\begin{eqnarray}
&&\bm{\mathcal {S}}(\bm{r},\bm{\Omega}, x)=
\sum_{k=-\infty}^{k=\infty}e^{ik\varphi}\,\,
{\tilde{\bm{\mathcal {S}}}^{(k)}}(\bm{r}, \theta, x),\nonumber\\
&&\bm{\mathcal {I}}(\bm{r},\bm{\Omega}, x)=
\sum_{k=-\infty}^{k=\infty}e^{ik\varphi}\,\,
{\tilde{\bm{\mathcal {I}}}^{(k)}}(\bm{r},\bm{\Omega}, x),\nonumber\\
\label{thm}
\end{eqnarray}
and that the Fourier coefficients ${\tilde{\bm{\mathcal {S}}}^{(k)}}$
and  ${\tilde{\bm{\mathcal {I}}}^{(k)}}$ satisfy a transfer equation
of the form
\begin{eqnarray}
&&-\frac{1}{\kappa_{\rm tot}(\bm{r}, x)}\bm{\Omega} \cdot
\bm{\nabla}{\tilde{\bm{\mathcal{I}}}}^{(k)}
(\bm{r}, \bm{\Omega}, x) = \nonumber \\
\!\!\!\!\!\!&&[{\tilde{\bm{\mathcal{I}}}}^{(k)}
(\bm{r}, \bm{\Omega}, x)-
{\tilde{\bm{\mathcal S}}}^{(k)}(\bm{r}, \theta, x)].
\label{f-rte}
\end{eqnarray}

\noindent
Proof: 
The proof is given for the general case of a frequency domain 
based PRD (approximation level II) that was derived by 
\citet{bom97a,bom97b}.
Since the angle-dependent PRD functions 
$r_{\rm II,III}(x,x',\bm{\Omega},\bm{\Omega'})$
are periodic functions of $\varphi$ with a period $2 \pi$, we 
can express them in terms of a Fourier series
\begin{eqnarray}
r_{\rm II,III}(x,x',\bm{\Omega},\bm{\Omega'})=
\sum_{k=-\infty}^{k=\infty}e^{ik\varphi}\,\,
{\tilde{r}}_{\rm II,III}^{(k)}(x,x',\theta,\bm{\Omega}'),
\label{Fourier-series-r23}
\end{eqnarray}
where the Fourier coefficients ${\tilde{r}}_{\rm II,III}^{(k)}$ are given by
\begin{eqnarray}
&&{\tilde{r}}_{\rm II,III}^{(k)}(x,x',\theta,\bm{\Omega'})\nonumber\\
&&=\int_0^{2\pi}\frac{d\,\varphi}{2\pi}\, e^{-ik\varphi}\,\,
r_{\rm II,III}(x,x',\bm{\Omega},\bm{\Omega'}).
\label{f-coefficients}
\end{eqnarray}
We let 
\begin{equation}
\bm{G}(\bm{r})=\sum_{k=-\infty}^{k=\infty}e^{ik\varphi}\,\,
{\tilde{\bm{G}}}^{(k)}(\bm{r}),
\end{equation}
where
\begin{equation}
{\tilde{\bm{G}}}^{(k)}(\bm{r})=
\int_0^{2\pi}\frac{d\,\varphi}{2\pi}\, e^{-ik\varphi}\,\,\bm{G}(\bm{r}).
\end{equation}
Note that 
\begin{eqnarray}
{\tilde{\bm{G}}}^{(k)}(\bm{r})=\begin{cases} 
\bm{G}(\bm{r}) 
& {\textrm{if}}\ k=0, 
\\
0 & {\textrm{if}}\ {{k \ne 0}}.
\end{cases}
\label{G}
\end{eqnarray}
We can write
\begin{equation}
\bm{\mathcal S}_C(\bm{r},x)=\sum_{k=-\infty}^{k=\infty}e^{ik\varphi}\,\,
{\tilde{\bm{\mathcal S}}}_C^{(k)}(\bm{r},x),
\label{sc-expansion}
\end{equation}
where
\begin{equation}
{\tilde{\bm{\mathcal S}}}_C^{(k)}(\bm{r},x)=\delta_{k0} \bm{\mathcal S}_C(\bm{r},x)
\label{f-sc}
\end{equation}
Substituting Equation~(\ref{Fourier-series-r23}) in Equation~(\ref{sl-reduced})
and using Equations~(\ref{f-sc}) and (\ref{stot-reduced}) we get 
\begin{eqnarray}
&&\bm{\mathcal {S}}(\bm{r},\bm{\Omega}, x)=
\sum_{k=-\infty}^{k=\infty}e^{ik\varphi}\,\,
{\tilde{\bm{\mathcal {S}}}^{(k)}}(\bm{r}, \theta, x),\nonumber\\
\label{s-expansion}
\end{eqnarray}
where 
\begin{eqnarray}
&&{\tilde{\bm{\mathcal {S}}}^{(k)}}(\bm{r}, \theta, x)=
p_x{\tilde{\bm{\mathcal {S}}}_l^{(k)}}(\bm{r}, \theta, x)
\nonumber\\
&&+(1-p_x){\tilde{\bm{\mathcal S}}}_C^{(k)}(\bm{r},x),
\label{f-stot}
\end{eqnarray}
with
\begin{eqnarray}
\!\!\!\!\!\!&&{\tilde{\bm{\mathcal {S}}}_l^{(k)}}(\bm{r}, \theta, x)=
{\tilde{\bm{G}}}^{(k)}(\bm{r})
+\frac{1}{\phi(x)} \int_{-\infty}^{+\infty} dx' \nonumber \\
\!\!\!\!\!\!&&
\times\oint\frac{d \bm{\Omega}'} {4 \pi} 
\hat{W}\Big\{\hat{M}_{\rm II}(\bm{B},x,x')
{\tilde{r}}^{(k)}_{\rm II}(x, x', \theta, \bm{\Omega'}) \nonumber \\
\!\!\!\!\!\!&&+\hat{M}_{\rm III}(\bm{B},x,x')
{\tilde{r}}^{(k)}_{\rm III}(x, x',\theta, \bm{\Omega'}) \Big\}
\nonumber \\
\!\!\!\!\!\!&&\times \hat{\Psi}(\bm{\Omega}') 
\bm{\mathcal{I}}(\bm{r}, \bm{\Omega}',x').
\label{f-sl}
\end{eqnarray}
Substituting Equation~(\ref{f-sl}) in Equation~(\ref{i-out-tau})
we get 
\begin{eqnarray}
&&\bm{\mathcal {I}}(\bm{r},\bm{\Omega}, x)=
\sum_{k=-\infty}^{k=\infty}e^{ik\varphi}\,\,
{\tilde{\bm{\mathcal {I}}}^{(k)}}(\bm{r}, \bm{\Omega}, x),\nonumber\\
\label{i-expansion}
\end{eqnarray}
where
\begin{eqnarray}
&&\!\!\!\!\!\!\!\!\!
{\tilde{\bm{\mathcal{I}}}}^{(k)}(\bm{r},\bm{\Omega}, x)=
{\tilde{\bm{\mathcal {I}}}}^{(k)}(\bm{r}_0, \bm{\Omega}, x)
e^{-\tau_{x, max}}\nonumber \\
&&\!\!\!\!\!\!\!\!\!+\int_{0}^{\tau_{x, max}} 
e^{-\tau'_{x}(\bm{r}')}
{\tilde{\bm{\mathcal {S}}}}^{(k)}
(\bm{r}', \theta, x)d\tau'_{x}(\bm{r}'),
\nonumber \\
\label{f-i-out-tau}
\end{eqnarray}
with
\begin{eqnarray}
{\tilde{\bm{\mathcal {I}}}}^{(k)}(\bm{r}_0, \bm{\Omega}, x)
=\delta_{k0}{\bm{\mathcal {I}}}(\bm{r}_0, \bm{\Omega}, x).
\end{eqnarray}
Here ${\tilde{\bm{\mathcal {S}}}}^{(k)}$ depends
only on $\bm{r}'$ but not the variable of integration $\tau'_{x}(\bm{r}')$
which is measured along a given ray determined by the
direction $\bm{\Omega}$.
Substituting Equation~(\ref{f-i-out-tau}) in 
Equation~(\ref{f-sl}) we obtain
\begin{eqnarray}
\!\!\!\!\!\!&&{\tilde{\bm{\mathcal {S}}}_l^{(k)}}(\bm{r}, \theta, x)=
{\tilde{\bm{G}}}^{(k)}(\bm{r})
+\frac{1}{\phi(x)} \int_{-\infty}^{+\infty} dx' \nonumber \\
\!\!\!\!\!\!&&
\times\oint\frac{d \bm{\Omega}'} {4 \pi} 
\hat{W}\Big\{\hat{M}_{\rm II}(\bm{B},x,x')
{\tilde{r}}^{(k)}_{\rm II}(x, x', \theta, \bm{\Omega'}) \nonumber \\
\!\!\!\!\!\!&&+\hat{M}_{\rm III}(\bm{B},x,x')
{\tilde{r}}^{(k)}_{\rm III}(x, x',\theta, \bm{\Omega'}) \Big\}
\hat{\Psi}(\bm{\Omega}') \nonumber\\
\!\!\!\!\!\!&&\sum_{k'=-\infty}^{k'=+\infty}e^{ik'\varphi'}
{\tilde{\bm{\mathcal{I}}}}^{(k')}(\bm{r}, \bm{\Omega}',x').
\label{f-sl-i}
\end{eqnarray}
Now from Equations~(\ref{s-expansion}) and 
(\ref{i-expansion}) and Equation~(\ref{rte-reduced})
it is straight forward to show that the Fourier coefficients 
${\tilde{\bm{\mathcal {S}}}^{(k)}}$
and  ${\tilde{\bm{\mathcal {I}}}^{(k)}}$ satisfy a transfer equation
of the form
\begin{eqnarray}
&&-\frac{1}{\kappa_{\rm tot}(\bm{r}, x)}\bm{\Omega} \cdot
\bm{\nabla}{\tilde{\bm{\mathcal{I}}}}^{(k)}
(\bm{r}, \bm{\Omega}, x) = \nonumber \\
\!\!\!\!\!\!&&[{\tilde{\bm{\mathcal{I}}}}^{(k)}
(\bm{r}, \bm{\Omega}, x)-
{\tilde{\bm{\mathcal S}}}^{(k)}(\bm{r}, \theta, x)].
\label{f-rte-p}
\end{eqnarray}
This proves the theorem.
Equation~(\ref{Fourier-series-r23}) represents the Fourier series expansion
of the angle-dependent redistribution functions 
$r_{\rm II, III}(x,x',\bm{\Omega},\bm{\Omega'})$. The expansion is with 
respect to the azimuth $\varphi$ of the scattered ray. In this respect
our expansion method differs from those used in \citet{dh88}, HF09, HF10 and
\citet{sametal11a}, all of whom perform expansion with respect to
$\varphi-\varphi'$, where $\varphi'$ is the incident ray azimuth. The
expansion used by these authors naturally leads to axisymmetry of the
Fourier components ${\tilde{\bm{\mathcal {I}}}^{(k)}}$, because
of the 1D planar geometry assumed by them. In a multi-D geometry
the expansion with respect to $\varphi-\varphi'$ does not provide
any advantage. In fact ${\tilde{\bm{\mathcal {I}}}^{(k)}}$
continue to depend on $\varphi$  due to finiteness of the co-ordinate
axes $X$ and/or $Y$ in multi-D geometry, under expansions either
with respect to $\varphi$ or $\varphi-\varphi'$. The Fourier
expansion of the ${\bm{\mathcal {S}}}$ in terms of
$\varphi$ (or $\varphi-\varphi'$) leads to axisymmetric 
${\tilde{\bm{\mathcal {S}}}^{(k)}}$ in 1D as well as multi-D geometries.
Thus both the approaches are equivalent. 

\subsection{SYMMETRY PROPERTIES OF THE IRREDUCIBLE 
FOURIER COEFFICIENTS}
From Equation~(\ref{f-coefficients}) it is easy to show 
that the components ${\tilde{r}}^{(k)}_{\rm II, III}$
satisfy the conjugation property
\begin{equation}
{\tilde{r}}^{(k)}_{\rm II, III}=\left
({\tilde{r}}^{(k)}_{\rm II, III}\right)^{*}.
\label{conj}
\end{equation}
In other words the real and imaginary parts 
of ${\tilde{r}}^{(k)}_{\rm II, III}$
are respectively symmetric and anti-symmetric 
about $k=0$.

Using Equation~(\ref{conj}) we can re-write
Equation~(\ref{Fourier-series-r23}) as
\begin{eqnarray}
&&r_{\rm II,III}(x,x',\bm{\Omega},\bm{\Omega'})=
{\tilde{r}}_{\rm II,III}^{(0)}(x,x',\theta,\bm{\Omega}')
\nonumber\\
&&+\sum_{k=1}^{k=\infty}\{e^{-ik\varphi}\,\,
{\tilde{r}}_{\rm II,III}^{(-k)}(x,x',\theta,\bm{\Omega}')
\nonumber\\
&&+e^{ik\varphi}\,\,{\tilde{r}}_{\rm II,III}^{(k)}(x,x',\theta,\bm{\Omega}')
\},
\label{Fourier-series-r23-p}
\end{eqnarray}
or
\begin{eqnarray}
&&r_{\rm II,III}(x,x',\bm{\Omega},\bm{\Omega'})=
\nonumber\\
&&\sum_{k=0}^{k=\infty}\,\, (2-\delta_{k0})
e^{ik\varphi}\,\,{\tilde{r}}_{\rm II,III}^{(k)}(x,x',\theta,\bm{\Omega}').
\label{Fourier-series-r23-pf}
\end{eqnarray}
In Equation~(\ref{Fourier-series-r23-pf}) the Fourier
series constitutes only the terms with $k \geq 0$. This is useful in 
practical applications.
With this simplification we can show, following the steps
similar to those given in Section~\ref{fourier-rte}, 
that Equation~(\ref{thm}) now becomes
\begin{eqnarray}
&&\bm{\mathcal {S}}(\bm{r},\bm{\Omega}, x)=
\sum_{k=0}^{k=\infty}e^{ik\varphi}\,\, (2-\delta_{k0})
{\tilde{\bm{\mathcal {S}}}^{(k)}}(\bm{r}, \theta, x),\nonumber\\
&&\bm{\mathcal {I}}(\bm{r},\bm{\Omega}, x)=
\sum_{k=0}^{k=\infty}e^{ik\varphi}\,\, (2-\delta_{k0})
{\tilde{\bm{\mathcal {I}}}^{(k)}}(\bm{r},\bm{\Omega}, x),\nonumber\\
\label{thm-new}
\end{eqnarray}
where 
\begin{eqnarray}
&&{\tilde{\bm{\mathcal {S}}}^{(k)}}(\bm{r}, \theta, x)=
p_x{\tilde{\bm{\mathcal {S}}}_l^{(k)}}(\bm{r}, \theta, x)
\nonumber\\
&&+(1-p_x){\tilde{\bm{\mathcal S}}}_C^{(k)}(\bm{r},x),
\label{f-stot-new}
\end{eqnarray}
with
\begin{equation}
{\tilde{\bm{\mathcal S}}}_C^{(k)}(\bm{r},x)=\delta_{k0} \bm{\mathcal S}_C(\bm{r},x)
\label{f-sc-new}
\end{equation}
and
\begin{eqnarray}
\!\!\!\!\!\!&&{\tilde{\bm{\mathcal {S}}}_l^{(k)}}(\bm{r}, \theta, x)=
{\tilde{\bm{G}}}^{(k)}(\bm{r})
+\frac{1}{\phi(x)} \int_{-\infty}^{+\infty} dx' \nonumber \\
\!\!\!\!\!\!&&
\times\oint\frac{d \bm{\Omega}'} {4 \pi} 
\hat{W}\Big\{\hat{M}_{\rm II}(\bm{B},x,x')
{\tilde{r}}^{(k)}_{\rm II}(x, x', \theta, \bm{\Omega'}) \nonumber \\
\!\!\!\!\!\!&&+\hat{M}_{\rm III}(\bm{B},x,x')
{\tilde{r}}^{(k)}_{\rm III}(x, x',\theta, \bm{\Omega'}) \Big\}
\hat{\Psi}(\bm{\Omega}') \nonumber\\
\!\!\!\!\!\!&&\sum_{k'=0}^{k'=+\infty}e^{ik'\varphi'}  (2-\delta_{k'0})
{\tilde{\bm{\mathcal{I}}}}^{(k')}(\bm{r}, \bm{\Omega}',x').
\label{f-sl-i-new}
\end{eqnarray}

\begin{figure*}
\centering
\includegraphics[scale=0.35]{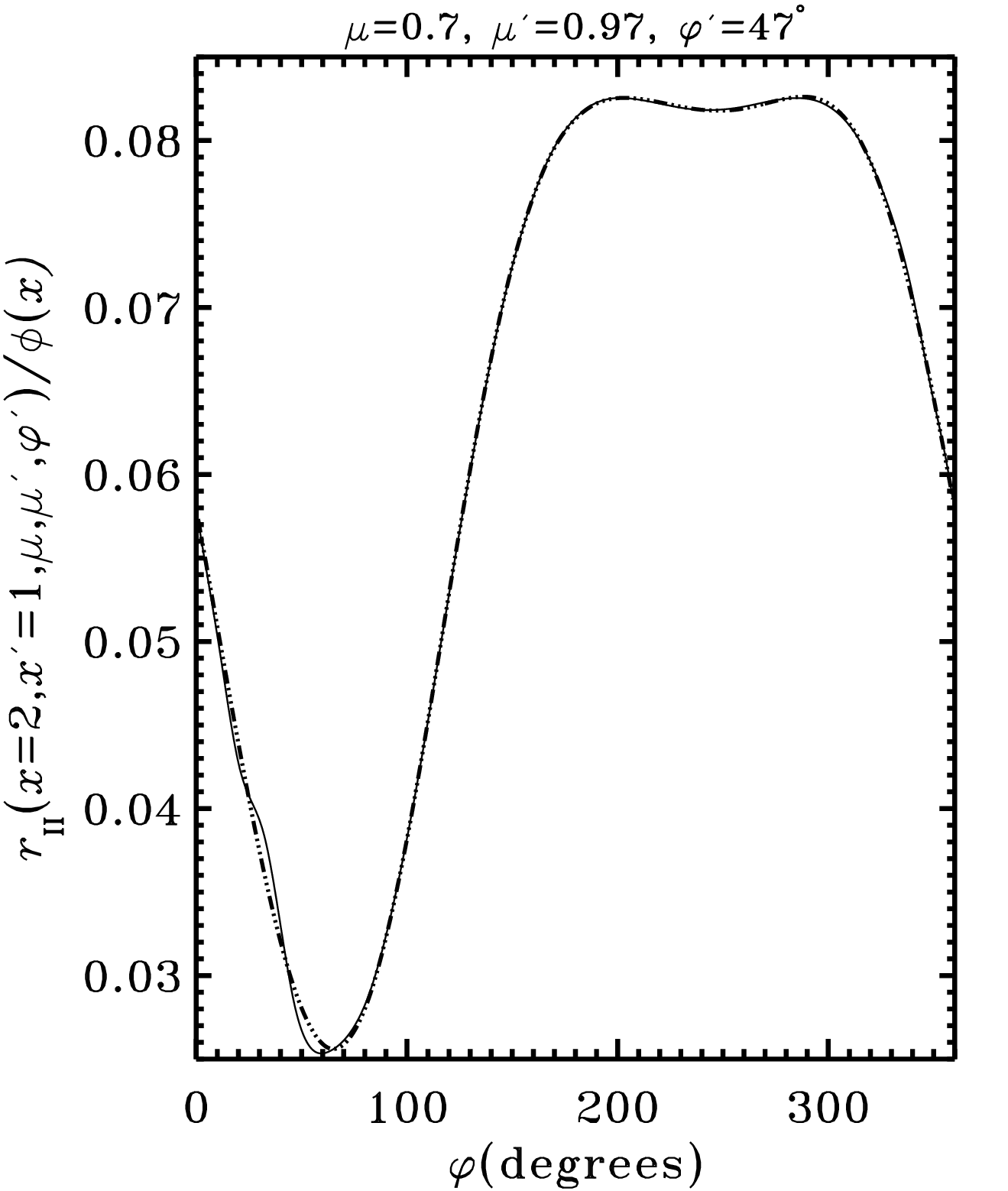}
\hspace{0.5cm}
\includegraphics[scale=0.35]{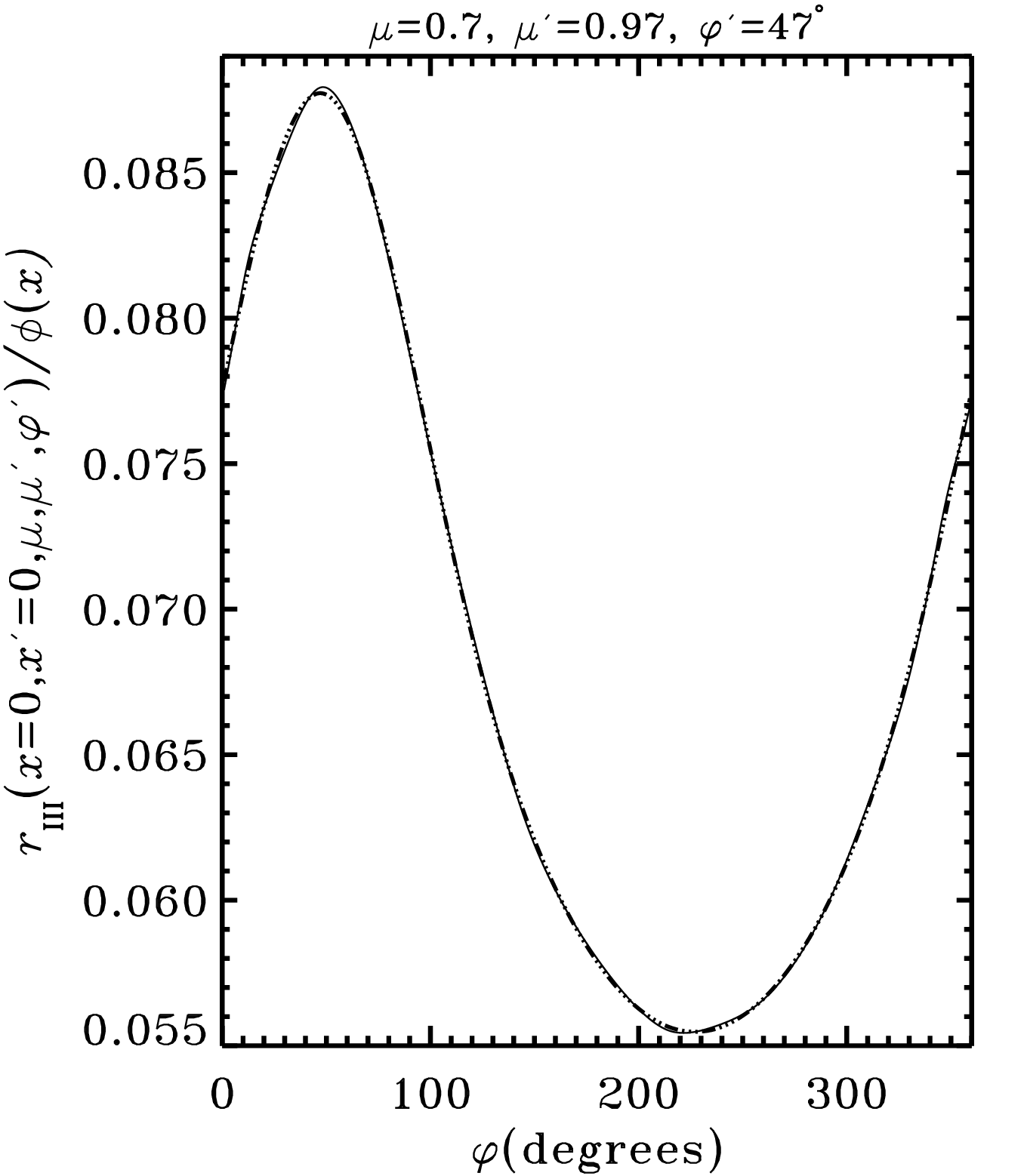}
\caption{Comparison of the exact (solid lines) and Fourier 
expansion (dash-triple-dotted lines) of 
$r_{\rm II}$ and $r_{\rm III}$ functions with 5 terms retained
in the series (Equation~(\ref{Fourier-series-r23-pf})).}
\label{fig-validate}
\end{figure*}

The components of ${\tilde{\bm{\mathcal {S}}}^{(k)}}$
and ${\tilde{\bm{\mathcal {I}}}^{(k)}}$ in general form countably
infinite sets{\footnote {If a set has an one-to-one correspondence with
the set of integers, it is called as a countably infinite set.}}.
We have verified that for practical applications it is sufficient to
work with five terms in the Fourier series ($k\in[0,+4]$).
Figure~\ref{fig-validate} shows a plot of the $r_{\rm II,III}$
functions computed using an exact method \citep[as in][]{knnetal02},
and those computed using Equation~(\ref{Fourier-series-r23-pf})
with $k \in [0,+4]$, namely keeping only the 5 dominant components
in the series expansion. A similar comparison of the exact and series
expansion methods for $r_{\rm II}$ is presented in \citet{dh88}, who
also show that 5 dominant components are sufficient to accurately represent
angle dependent $r_{\rm II}$ function.

\begin{figure*}
\centering
\includegraphics[scale=0.35]{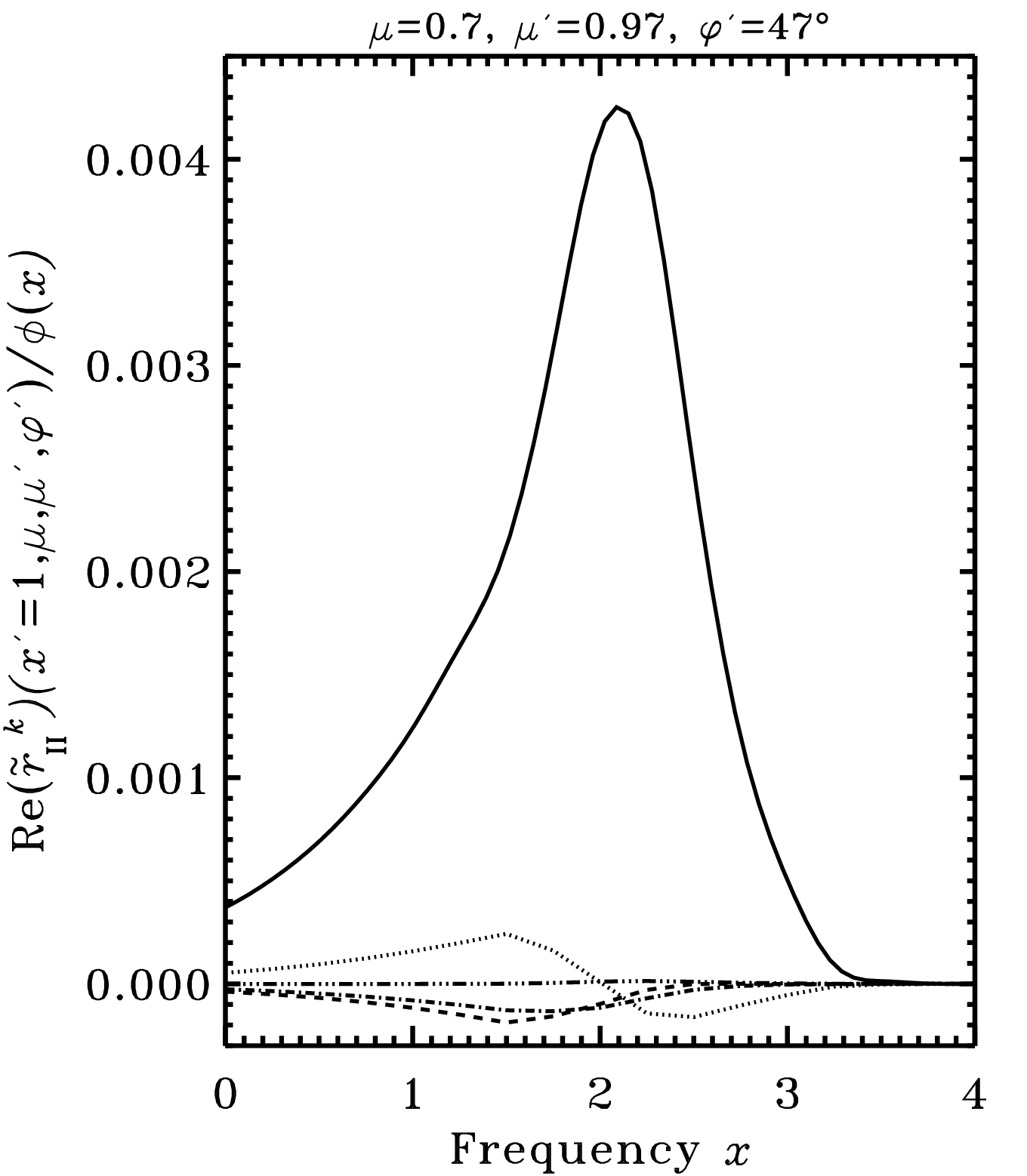}
\hspace{0.5cm}
\includegraphics[scale=0.35]{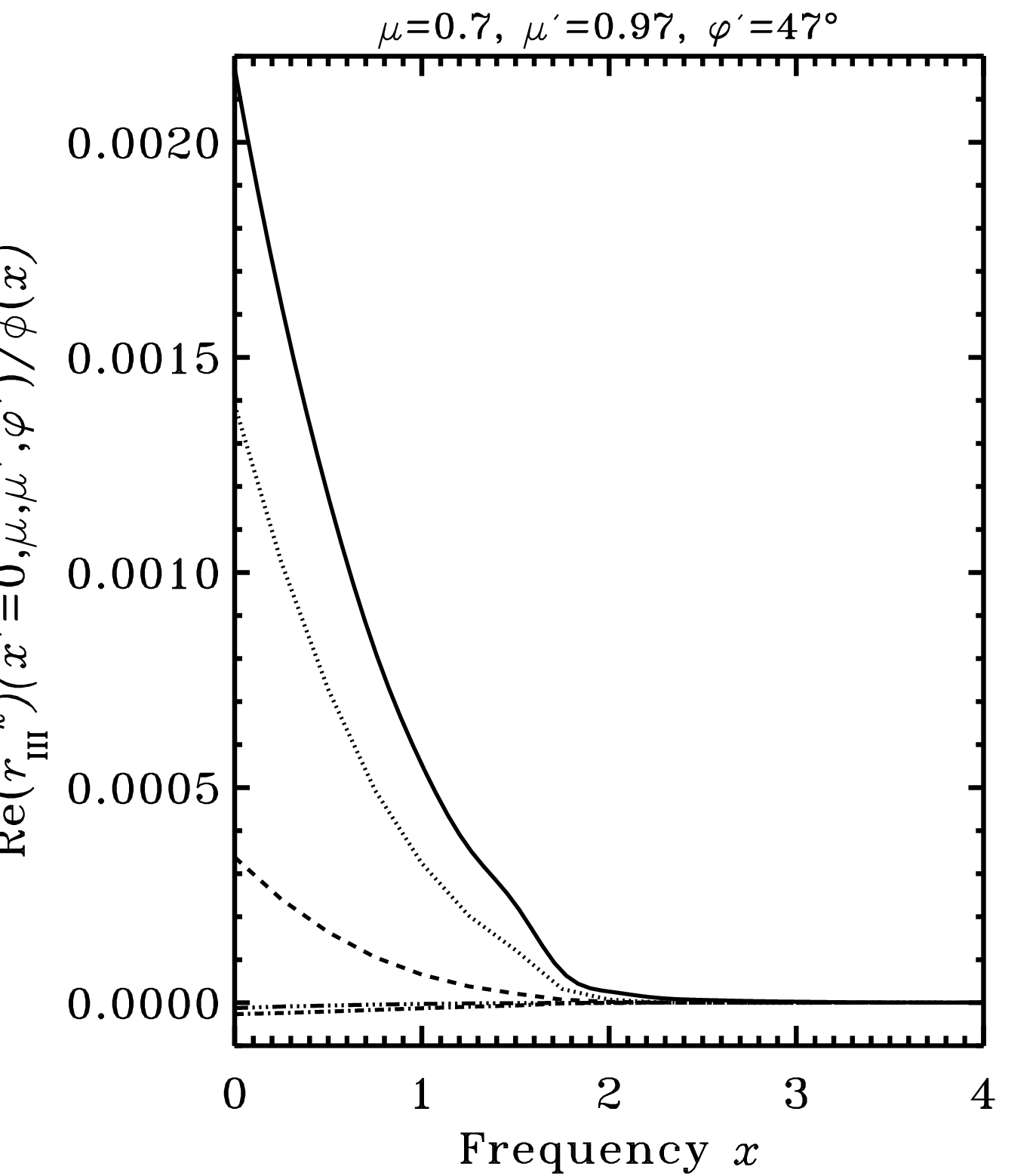}
\caption{Frequency dependence of real parts of 
${\tilde{r}}^{(k)}_{\rm II, III}(x,x',\theta,\bm{\Omega'})$.
Solid, dotted, dashed, dot-dashed and dash-triple-dotted
lines respectively correspond to $k=0$, $k=1$, $k=2$,
$k=3$ and $k=4$.
}
\label{fig-rtildareal}
\end{figure*}

\begin{figure*}
\centering
\includegraphics[scale=0.35]{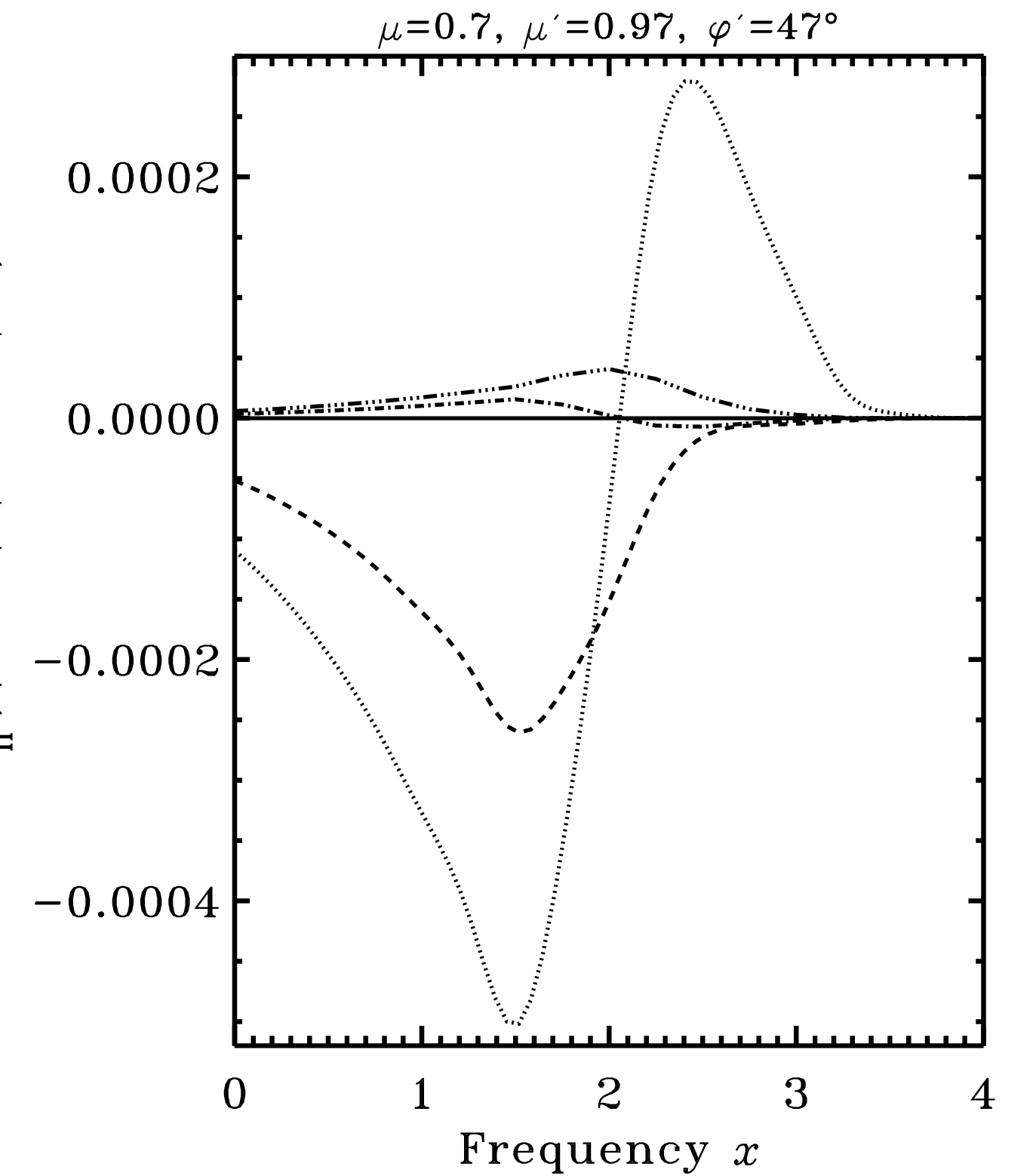}
\hspace{0.8cm}
\includegraphics[scale=0.35]{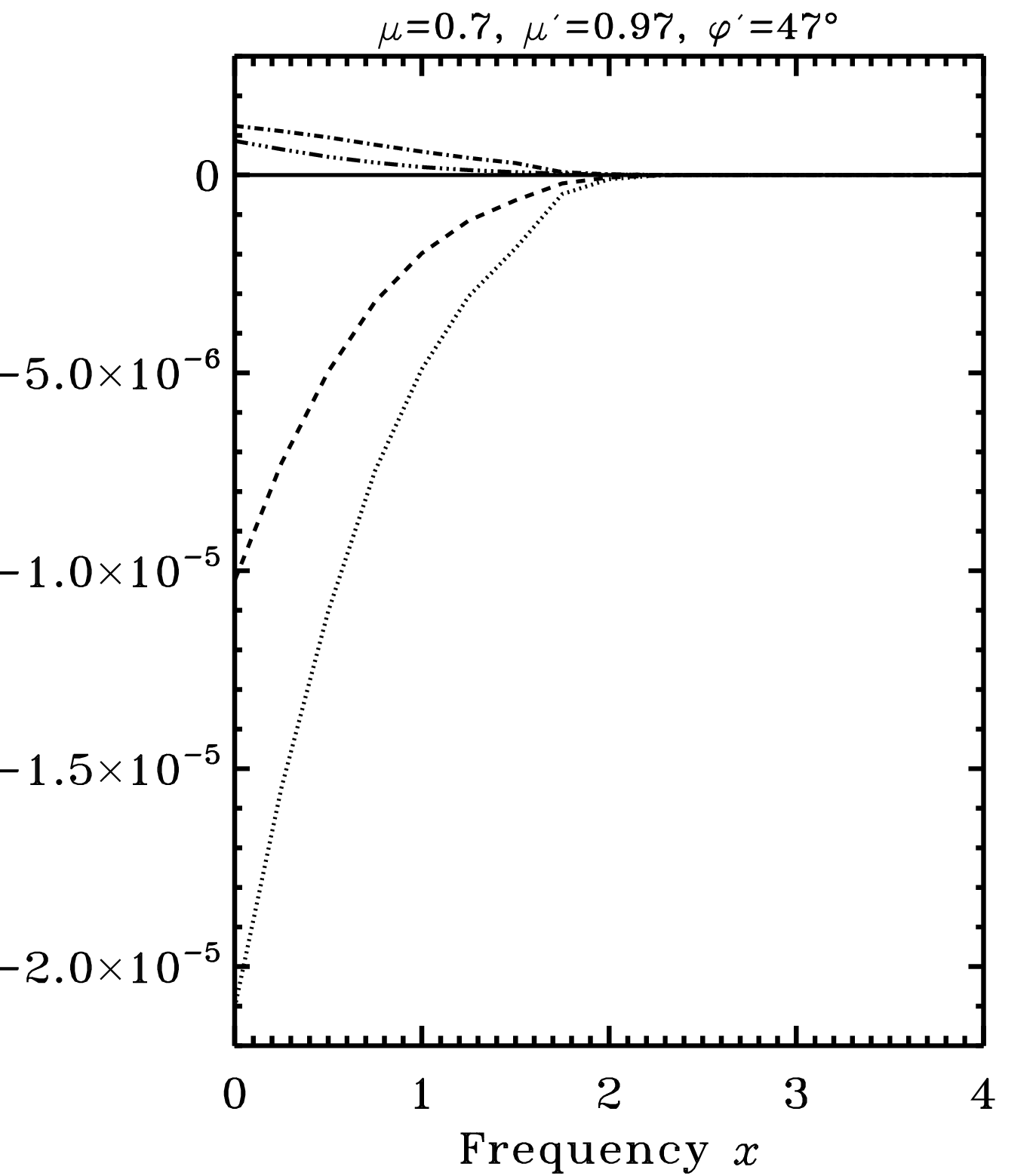}
\caption{Frequency dependence of imaginary parts of 
${\tilde{r}}^{(k)}_{\rm II, III}(x,x',\theta,\bm{\Omega'})$.
Solid, dotted, dashed, dot-dashed and dash-triple-dotted
lines respectively correspond to $k=0$, $k=1$, $k=2$,
$k=3$ and $k=4$.
}
\label{fig-rtildaimag}
\end{figure*}

In Figures~\ref{fig-rtildareal} and \ref{fig-rtildaimag} 
we study the frequency dependence of 
the real and imaginary parts of 
${\tilde{r}}^{(k)}_{\rm II, III}(x,x',\theta,\bm{\Omega'})$ 
for a given incident 
frequency point ($x'=2$ for ${\tilde{r}}^{(k)}_{\rm II}$ and 
$x'=0$ for ${\tilde{r}}^{(k)}_{\rm III}$). 
We show the behaviour of five ($k=0, 1, 2, 3, 4$)
Fourier components. 
Note that ${\tilde{r}}^{(0)}_{\rm II, III}$ are real quantities.

Equations~(\ref{f-rte-p}) and (\ref{f-i-out-tau}) together with 
Equations~(\ref{f-sl-i-new}), (\ref{f-sc-new}) and (\ref{f-stot-new}) 
can be solved using an iterative method. In a subsequent paper we develop a fast
iterative method (pre-BiCG-STAB) and present
the solutions of polarized RT in multi-D geometry including Hanle effect
with angle-dependent PRD. 

After solving for ${\tilde{\bm{\mathcal {S}}}^{(k)}}$ and
${\tilde{\bm{\mathcal {I}}}^{(k)}}$ we can construct $\bm{\mathcal S}$
and $\bm{\mathcal I}$ using the Equation~(\ref{thm-new}). Since 
$\bm{\mathcal S}$ and $\bm{\mathcal I}$
are real quantities these expansions reduce to
the following simpler forms.
\begin{eqnarray}
&&\bm{\mathcal S}(\bm{r}, \bm{\Omega}, x)=\nonumber\\
&&\sum_{k=0}^{k=\infty}(2-\delta_{k0})
\Big\{\cos(k\varphi){\mathcal Re}\{
{\tilde{\bm{\mathcal {S}}}^{(k)}}(\bm{r}, \theta, x)\}\nonumber\\
&&-\sin(k\varphi){\mathcal Im}\{{\tilde{\bm{\mathcal {S}}}^{(k)}}
(\bm{r}, \theta, x)\} \Big\},\nonumber\\
\label{transform-1}
\end{eqnarray}
and
\begin{eqnarray}
&&\bm{\mathcal I}(\bm{r}, \bm{\Omega}, x)=\nonumber\\
&&\sum_{k=0}^{k=\infty}(2-\delta_{k0})\Big\{\cos(k\varphi){\mathcal Re}\{
{\tilde{\bm{\mathcal {I}}}^{(k)}}(\bm{r}, \bm{\Omega}, x)\}\nonumber\\
&&-\sin(k\varphi){\mathcal Im}\{{\tilde{\bm{\mathcal {I}}}^{(k)}}
(\bm{r}, \bm{\Omega}, x)\} \Big\},\nonumber\\
\label{transform-2}
\end{eqnarray}
where
$\bm{\mathcal S}=(S^0_0$,
$S^2_0$, $S^{2,{\rm x}}_1$, $S^{2,{\rm y}}_1$,
$S^{2,{\rm x}}_2$, $S^{2,{\rm y}}_2)^T$ and 
$\bm{\mathcal I}=(I^0_0$, $I^2_0$, $I^{2,{\rm x}}_1$, $I^{2,{\rm y}}_1$,
$I^{2,{\rm x}}_2$, $I^{2,{\rm y}}_2)^T$.

Once we obtain $\bm{\mathcal S}$ and $\bm{\mathcal I}$, the Stokes
source vector and Stokes intensity vector can be deduced using the 
following formulae (see Appendix B of HF07 and also Paper I).
\begin{eqnarray}
&&I(\bm{r}, \bm{\Omega}, x) = I^0_0 +
\frac{1}{2 \sqrt{2}} (3 \cos^2\theta -1) I^2_0 \nonumber \\
&&-\sqrt{3} \cos \theta \sin \theta (I^{2, {\rm x}}_1 
\cos \varphi-I^{2, {\rm y}}_1 \sin \varphi) \nonumber \\ 
&&+ \frac{\sqrt{3}}{2} (1-\cos^2\theta)
(I^{2, {\rm x}}_2 \cos 2 \varphi-I^{2, {\rm y}}_2 \sin 2 \varphi), \nonumber \\
\label{transform-old1}
\end{eqnarray}
\begin{eqnarray}
&&Q(\bm{r}, \bm{\Omega}, x)= -\frac{3}{2 \sqrt{2}} 
(1- \cos^2\theta) I^2_0 \nonumber \\
&&-\sqrt{3} \cos \theta \sin \theta (I^{2, {\rm x}}_1 
\cos \varphi-I^{2, {\rm y}}_1 \sin \varphi) \nonumber \\ 
&&-\frac{\sqrt{3}}{2} (1+\cos^2\theta)
(I^{2, {\rm x}}_2 \cos 2 \varphi-I^{2, {\rm y}}_2 \sin 2 \varphi),\nonumber \\
\label{transform-old2}
\end{eqnarray}

\begin{eqnarray}
&&U(\bm{r}, \bm{\Omega}, x) = \sqrt{3} \sin \theta
(I^{2, {\rm x}}_1 \sin \varphi+I^{2, {\rm y}}_1 \cos \varphi) \nonumber \\ 
&&+ \sqrt{3} \cos \theta 
(I^{2, {\rm x}}_2 \sin 2 \varphi+I^{2, {\rm y}}_2 \cos 2 \varphi).
\label{transform-old3}
\end{eqnarray}
The quantities $I^0_0$, $I^2_0$, $I^{2,{\rm x}}_1$, $I^{2,{\rm y}}_1$,
$I^{2,{\rm x}}_2$, $I^{2,{\rm y}}_2$ also depend on 
$\bm{r}, \bm{\Omega}, x$. Similar formulae can also
be used to deduced $\bm{S}$ from $\bm{\mathcal{S}}$.

\section{NUMERICAL CONSIDERATIONS}
\label{numerics}
The proposed Fourier series expansion (or Fourier decomposition) 
technique to solve multi-D
RT problems with angle-dependent PRD functions essentially transforms
the given problem in $(\theta,\varphi)$ space (see Section~\ref{reduced-rte})
into $(\theta,k)$ space (see Section \ref{fourier-rte}). Let $n_{\varphi}$ 
denote the number of azimuths ($\varphi$) used in the computations and 
$n_k$ the maximum number of terms retained in the Fourier series expansions. 
In $(\theta,\varphi)$ space the source terms ${\bm{\mathcal {S}}}$ 
depend on $n_{\varphi}$ whereas in $(\theta,k)$ space the source 
terms ${\tilde{\bm{\mathcal {S}}}^{(k)}}$ depend on $n_k$.
In Figure~\ref{fig-validate} we have demonstrated that it is sufficient
to work with $n_k=5$ (i.e., $k\in[0, 4]$), whereas for 2D RT problems it
is necessary to use $n_{\varphi}$=8, 16, 24 or 32 depending upon
the accuracy requirements of the problem. Since $n_{k}$ is
always smaller than $n_{\varphi}$, the computational cost is reduced 
when we work in the $(\theta,k)$ space. 

In addition to the computation of 
$r_{\rm II,III}(x,x',\bm{\Omega},\bm{\Omega'})$ functions, we need to compute
${\tilde{r}}^{(k)}_{\rm II, III}(x,x',\theta,\bm{\Omega'})$ in 
the $(\theta,k)$ space. This additional computation does not require
much of CPU time. Moreover, if we can fix the number of angles and frequency
points to be used in the computations, it is sufficient to compute these 
functions only once, which can be written to a file. In subsequent transfer
computations this data can be simply read from the archival file.

To demonstrate these advantages, we have compared the CPU time requirements
for the two methods, namely the one which uses $(\theta,\varphi)$ space and
the other which uses the $(\theta,k)$ space. Both the approaches use the 
Pre-BiCG-STAB as the iterative method to solve the 2D transfer problem.
We find that with $n_{\varphi}=32$, the CPU time required to solve a 
given problem in $(\theta,k)$ space is 7 times lesser than that required 
to solve the same problem in $(\theta,\varphi)$ space. For practical problems
requiring more number of azimuthal angles, the advantages of using Fourier
decomposition technique is much larger.

\begin{figure*}
\centering
\includegraphics[scale=0.6]{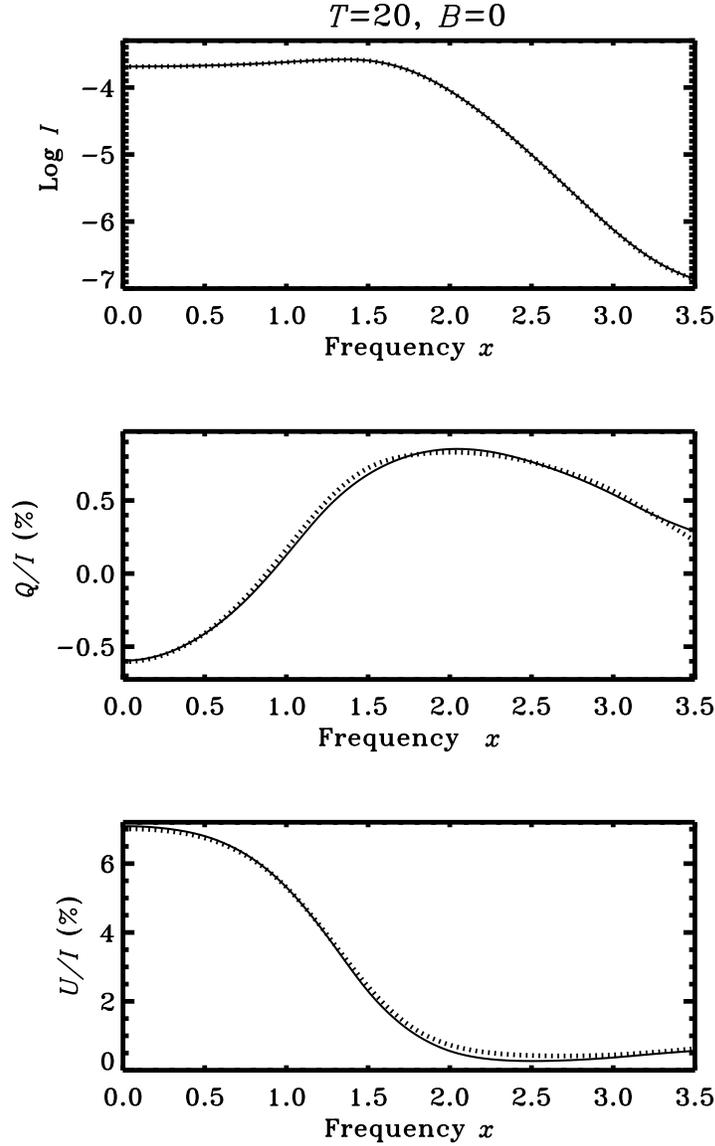}
\caption{Emergent, spatially averaged $I, Q/I, U/I$ profiles
computed for a test 2D RT problem with angle dependent PRD
using two methods, namely the one which uses $(\theta,k)$ space (solid lines)
and the other which uses the  $(\theta,\varphi)$ space (dotted lines)
Both the approaches use the Pre-BiCG-STAB as the iterative method. 
Both methods produce nearly identical results, proving the correctness 
of the proposed Fourier decomposition technique for angle-dependent
PRD problems in multi-D RT. The results are 
plotted for $\mu=0.1$ and $\varphi=27^{\circ}$. The details and other model 
parameters are given in Section~\ref{numerics}.
}
\label{fig-2dresult}
\end{figure*}

To demonstrate the correctness of the proposed 
Fourier decomposition technique for multi-D transfer, we consider
a test RT problem in 2D medium. A complete study of the solutions
of 2D RT problems with angle-dependent PRD will be taken up in a forthcoming
paper. Figure~\ref{fig-2dresult} shows the emergent, spatially averaged Stokes
profiles formed in a 2D media, computed using the two methods mentioned above.
The model parameters are:\, the total optical thickness in two directions namely
$T_Y=T_Z=T=20$, the elastic and inelastic collision rates respectively are 
$\Gamma_E/\Gamma_R=10^{-4}$, $\Gamma_I/\Gamma_R=10^{-4}$, the damping parameter
of the Voigt profile is $a=2 \times 10^{-3}$.
We consider the pure line case ($\kappa_c=0$). 
The internal thermal sources are taken as constant (the Planck 
function $B_{\nu}=1$). The medium
is assumed to be self-emitting (no incident radiation on the
boundaries). We consider the case of zero magnetic field.
The branching ratios for this choice of model
parameters are $(\alpha,\beta^{(0)},\beta^{(2)}) = (1, 1, 1)$.
These branching ratios correspond to a PRD scattering
that uses only ${\tilde{r}}^{(k)}_{\rm II}(x,x',\theta,\bm{\Omega'})$ 
function. We use a logarithmic frequency grid with $x_{\rm max}=3.5$ and 
a logarithmic depth grid in $Y$ and $Z$ directions of the 2D medium. 
We have used a 3-point Gaussian $\mu$-quadrature and a 32-point
Gaussian $\varphi$-quadrature. In Figure~\ref{fig-2dresult} we show the
results computed at $\mu=0.1$ and $\varphi=27^{\circ}$.
The fact that both the methods give nearly identical results proves the 
correctness of the proposed Fourier decomposition technique for multi-D RT.

\section{CONCLUSIONS}
\label{conclusions}
In this paper we formulate polarized RT equation in multi-D media
that includes angle-dependent PRD and Hanle effect. We propose
a method of decomposition of the Stokes source vector 
and Stokes intensity vector in terms of irreducible Fourier
components $\tilde{\bm{\mathcal{S}}}^{(k)}$ and 
$\tilde{\bm{\mathcal{I}}}^{(k)}$, using a 
combination of the decomposition of the 
scattering phase matrices in terms of irreducible spherical tensors
${\mathcal T}^K_Q$ and the Fourier series expansions of 
angle-dependent PRD functions. We also establish that the
irreducible Fourier components $\tilde{\bm{\mathcal{S}}}^{(k)}$
and $\tilde{\bm{\mathcal{I}}}^{(k)}$ satisfy a
simple transfer equation, which can be solved by any iterative method
like an Approximate Lambda Iteration (ALI) or a Biconjugate-Gradient type
projection method.

\acknowledgments
\noindent
We would like to thank Prof. H. Frisch for helpful comments/suggestions 
which helped to improve the manuscript. We thank Dr. Sampoorna for useful
discussions.

\begin{table*}
\tiny
\caption{Table describing the evolution of ideas in the past
three decades to simplify the difficult problem of formulating/solving
the polarized line transfer equation.
In the text of the table, we use the following
abbreviations. RTE:\,radiative transfer equation;\, AA:\,angle-averaged;\,
AD:\,angle-dependent;\,PM:\,phase matrix;\,RM:\,redistribution matrix,
CRD:\,complete frequency redistribution;\,PRD: partial frequency
redistribution.}
\label{table_1}
\vspace{0.6cm}
\begin{tabular}{lrr}
\tableline\tableline
Milestones&$\bm{B}=0$ (Rayleigh scattering)&$\bm{B}\ne0$ (Hanle effect)\\
\tableline\tableline
(1) Formulation of PM & \citet{chandra46} & \citet{jos78}\\
in Stokes vector formulation &\citet{hm47}  &\\
\tableline
(2) Stokes vector RTE : 1D/CRD & \citet{rees78} & \citet{mf91}\\
&  & \citet{knnetal02}\\
\tableline
(3) Stokes vector RTE : multi-D/CRD & \citet{fpetal99}&  \\
\tableline
(4) Stokes vector RTE : & \citet{reessalib82} : AA & \citet{mf91} : AA\\
1D/PRD                               & \citet{dumetal77} : AD &  \citet{knnetal02} : AA/AD\\
                               & \citet{knn86} : AA & \\
                               & \citet{mf87} : AA/AD & \\
\tableline
(5) PM decomposition & Landi Degl'Innocenti \& & Landi Degl'Innocenti \&\\
in terms of $\mathcal{T}^K_Q$ & Landi Degl'Innocenti (1988) & Landi Degl'Innocenti (1988)\\
\tableline
(6) Irreducible Stokes source&Landi Degl'Innocenti&Landi Degl'Innocenti\\
vector in Stokes vector RTE: 1D/CRD & et al. (1987)& et al. (1987)\\
\tableline
(7) Irreducible Stokes source& Manso Sainz \& & Manso Sainz \&\\
vector in Stokes vector RTE: & Trujillo Bueno (1999)& Trujillo
Bueno (1999)\\
multi-D/CRD &\citet{ditt99}&\citet{ditt99}\\
\tableline
(8) Irreducible Stokes& \citet{hf07} & \citet{hf07} \\
vector RTE: 1D/CRD & & \\
\tableline
(9) Formulation of polarized RM : &\citet{osc72}&\citet{osc73}\\
&\citet{dh88}&\citet{bom97a,bom97b}\\
\tableline
(10) RTE with RM : 1D/AA & \citet{mf91} & \citet{knnetal02}\\
& \citet{knn94} & \\
\tableline
(11) RTE with RM : & \citet{anuknn11a} & \citet{anuknn11c} \\
multi-D/AA & \citet{anuetal11b} & \\
\tableline
(12) RTE with RM : 1D/AD & \citet{mf87}&\citet{knnetal02}\\
& \citet{knnetal02}&\citet{sametal08}\\
\tableline
(13) Fourier decomposition of & \citet{hf09,hf10} & \citet{hf09}\\
AD PRD functions: 1D     & & \\
\tableline
(14) RTE with RM based on & \citet{sametal11a} & \citet{sam11b} \\
Fourier expansions of & & \citet{msknn11c}\\
AD PRD functions : 1D& &\\
\tableline
(15) a. RTE with RM : multi-D/AD & &\\
b. Fourier expansion of & Present paper and & Present paper and   \\
AD PRD functions: multi-D & Forthcoming paper& Forthcoming paper \\
c. RTE with RM based on &  & \\
Fourier expansions of & & \\
AD PRD functions : multi-D & &\\
\tableline
\tableline
\end{tabular}
\end{table*}

\begin{thebibliography}{}
\bibitem[Adam(1990)]{adam90}
Adam, J. 1990, \aap, 240, 541
\bibitem[Anusha \& Nagendra (2011a)]{anuknn11a}
Anusha, L. S., \& Nagendra, K. N. 2011a, \apj, 726, 6 (Paper I)
\bibitem[Anusha et al.(2011b)]{anuetal11b}
Anusha, L. S., Nagendra, K. N., \& Paletou, F. \apj, 2011b, 726, 96 (Paper II)
\bibitem[Anusha \& Nagendra (2011c)]{anuknn11c}
Anusha, L. S., \& Nagendra, K. N. 2011c (submitted), (Paper III)
\bibitem[Bommier(1997a)]{bom97a} 
Bommier, V. 1997, \aap, 328, 706
\bibitem[Bommier(1997b)]{bom97b}
Bommier, V. 1997b, \aap, 328, 726
\bibitem[Chandrasekhar(1946)]{chandra46}
Chandrasekhar, S. 1946, \apj, 103, 351
\bibitem[Chandrasekhar(1960)]{chandra60}
Chandrasekhar, S. 1960, Radiative Transfer (New York: Dover)
\bibitem[Dittmann(1999)]{ditt99}
Dittmann, O. J. 1999, in Solar Polarization,
ed. K. N. Nagendra \& J. O. Stenflo (Boston: Kluwer), 201
\bibitem[Domke \& Hubeny(1988)]{dh88}
Domke, H., \& Hubeny, I. 1988, \apj, 334, 527
\bibitem[Dumont et al.(1977)]{dumetal77}
Dumont, S., Omont, A., Pecker, J. C., \& Rees, D. E. 1977, \aap, 54, 675
\bibitem[Faurobert(1987)]{mf87}
Faurobert, M. 1987, \aap, 178, 269
\bibitem[Faurobert(1988)]{mf88}
Faurobert, M. 1988, \aap, 194, 268
\bibitem[Faurobert-Scholl(1991)]{mf91}
Faurobert-Scholl, M. 1991, \aap, 246, 469
\bibitem[Frisch(2007)]{hf07}
Frisch, H. 2007, \aap, 476, 665 (HF07)
\bibitem[Frisch(2009)]{hf09}
Frisch, H. 2009, in ASP Conf. Ser. 405, Solar Polarization 5, ed.
S. V. Berdyugina, K. N. Nagendra \& R. Ramelli (San Francisco: ASP), 
87 (HF09)
\bibitem[Frisch(2010)]{hf10}
Frisch, H. 2010, \aap, 522, A41 
\bibitem[Hamilton(1947)]{hm47}
Hamilton, D. R. 1947, \apj, 106, 457
\bibitem[Hummer(1962)]{hum62}
Hummer, D.~G. 1962, \mnras, 125, 21
\bibitem[Landi Degl'Innocenti \& Landi Degl'Innocenti(1988)]{ll88}
Landi Degl'Innocenti, M., \& Landi Degl'Innocenti, E. 1988, \aap, 192, 374
\bibitem[Landi Degl'Innocenti et al.(1987)]{landietal87}
Landi Degl'Innocenti, E., Bommier, V., \& Sahal-Br\'echot, S. 
1987, \aap, 186, 335
\bibitem[Landi Degl'Innocenti \& Landolfi(2004)]{ll04} 
Landi Degl'Innocenti, E., \& Landolfi, M. 2004, 
Polarization in Spectral Lines (Dordrecht: Kluwer) 
\bibitem[McKenna(1985)]{mck85}
McKenna, S. J. 1985, Ap\&SS, 108, 31
\bibitem[Manso Sainz \& Trujillo Bueno(1999)]{msjtb99}
Manso Sainz, R., \& Trujillo Bueno, J. 1999, in Solar Polarization,
ed. K. N. Nagendra \& J. O. Stenflo (Boston: Kluwer), 143
\bibitem[Mihalas et al.(1978)]{mam78}
Mihalas, D., Auer, L. H., \& Mihalas, B. R. 1978, \apj, 220, 1001
\bibitem[Nagendra(1986)]{knn86}
Nagendra, K.~N. 1986, PhD Thesis, Bangalore University
\bibitem[Nagendra(1994)]{knn94}
Nagendra, K.~N. 1994, \apj, 432, 274
\bibitem[Nagendra et al.(2002)]{knnetal02}
Nagendra, K.~N., Frisch, H., \& Faurobert, M. 2002, \aap, 395, 305
\bibitem[Nagendra et al.(2003)]{knnetal03}
Nagendra, K.~N., Frisch, H., \& Fluri, D. M. 2003, in ASP Conf. Ser. 307, 
Solar Polarization 3, ed. J. Trujillo Bueno \& J. S\'anchez Almeida 
(San Francisco: ASP), 227
\bibitem[Nagendra \& Sampoorna (2011)]{msknn11c}
Nagendra, K. N., \& Sampoorna, M. 2011, \aap, (submitted)
\bibitem[Omont et al.(1972)]{osc72}
Omont, A., Smith E. W., \& Cooper, J. 1972, \apj, 175, 185
\bibitem[Omont et al.(1973)]{osc73}
Omont, A., Smith E. W., \& Cooper, J. 1973, \apj, 182, 283
\bibitem[Paletou et al.(1999)]{fpetal99}
Paletou, F., Bommier, V., \& Faurobert-Scholl, M. 1999, in Solar Polarization,
ed. K. N. Nagendra \& J. O. Stenflo (Boston: Kluwer), 189
\bibitem[Pomraning(1973)]{pom73}
Pomraning, G. C. 1973, The equations of radiation hydrodynamics
(Oxford: Pergamon Press)
\bibitem[Rees(1978)]{rees78}
Rees, D. E., 1978, PASJ, 30, 455
\bibitem[Rees \& Saliba(1982)]{reessalib82}
Rees, D. E., \& Saliba, G. J. 1982, \aap, 115, 1
\bibitem[Sampoorna (2011b)]{sam11b}
Sampoorna, M. 2011, \aap, (in press) 
\bibitem[Sampoorna et al.(2008)]{sametal08}
Sampoorna, M., Nagendra, K. N., \& Stenflo, J. O. 2008, \apj, 679, 889
\bibitem[Sampoorna et al.(2011a)]{sametal11a}
Sampoorna, M., Nagendra, K. N., \& Frisch, H. 2011a, \aap, 527, A89
\bibitem[Stenflo(1978)]{jos78}
Stenflo, J. O. 1978, \aap, 66, 241
\end{thebibliography}
\end{document}